# Is there Ballistic Conductance Quantization in Real Life Metals Nanocontacts?

N. García, Ming Bai, Yonghua Lu and M. Muñoz

Laboratorio de Física de Sistemas Pequeños y Nanotecnología, Consejo Superior de investigaciones científicas (CSIC), Madrid 28006, Spain

A. P. Levanyuk,

Fisintec Innovación Tecnológica, Miraflores 65, Alcobendas, Madrid 28100, Spain

Theory and a vast set of experimental work in metals, since a century, appear to show that the mean free path of conduction electrons in a real metal is about the **min** (*bulk mean free path, smallest transversal size of the metal*), a result that was already proposed by J. J. Thomson in 1901. This establishes, as discussed in this work, serious difficulties to justify conductance quantization and ballistic transport in atomic/nanocontacts or nanoconstrictions of real life metals. The ohmic resistance of the leads proves to be as important as the ballistic one of the constriction.

PACS numbers: 73.63.Rt, 73.23.Ad

In 1901, J. J. Thomson [1] was the first to suggest that the source of the high resistivity $\rho$ of very thin films of metals lay in the limitation of the mean free path of the electrons due to non specular scattering at the surface of the films. With the advent of low temperatures and film growth techniques, Lovell [2] and Andrew [3] realized fantastic experiments proven the suggestion of Thomson. These experiments were done in Sn films and Na wires. Also theoretical work was done by Fuchs [4], Sondheimer [5] and Chambers [6] on thin films using the Boltzmann equation and taking into account the specularity in the $p$ parameter for the surface scattering to explain the data. Dingle [7] and MacDonald and Sarginson [8] did the same kind of work for circular and square sectioned wires respectively. These results are also described in a book on metals edited by Ziman [9].

The resume up for different values of $\kappa$ is as follows [5]:

$$\frac{\rho}{\rho_0} = \frac{4}{3}\frac{1-p}{1+p}\frac{1}{\kappa \log(1/\kappa)} \quad \kappa \ll 1 \quad (1)$$

for thin films;

$$\frac{\rho}{\rho_0} = \frac{1-p}{1+p}\frac{1}{\kappa} \quad \kappa \ll 1 \tag{2}$$

for circular wires

$$\frac{\rho}{\rho_0} = \frac{1-p}{1+p}\frac{0.897}{\kappa} \quad \kappa \ll 1 \tag{3}$$

square wires.

Not necessary to mention that the resistivity in thin films has been a field of very high activity for many materials, in the late years, of which we mentioned few [10-15]. The classical theory shows that for the ratio $\kappa=a/l \ll 1$, where $a$ is the smallest transversal size of the object at hand (the thickness for films or the diameter for wires) and $l$ is the mean free path in the bulk, $\rho$ behaves approximately as $\rho_0/\kappa$ *($\rho_0$ is the bulk resistivity due to collisions)*. This means that the effective mean free path of the object $l_e$ is controlled by the transversal size $a$, and has little resemblance with that of the bulk material $l$. **An important result that puts under serious considerations the concept of ballistic transport in small or nanometer real metallic objects at any temperature because $l_e$ is never much larger than $a$; i.e the condition required for non-ballistic transport .**

Fig.1a illustrates the behaviour described above. Further developments by Mayadas and Shatzkes [16] introduced the fact that grain boundary scattering in polycrystalline film will increase the resistivity in addition to surface scattering. The thinner is the film; more accentuated it is, because the grain boundaries reduce in size. In Fig. 1a we present a summed up set of modern data on thin film's resistivities for several metals as well as the fit using the Fuchs-Sondheimer theory for the best values of $p$ as well as the parameter $R_{MS}$ of the Mayadas and Shatzkes theory.

All the previous theory was done using the classical Boltzmann's equation. However, quantum calculations showed that the effect of increasing resistivity for very thin films $\kappa<0.2$ is more dramatic than the prediction of classical theory and there is a crossover from $\rho$ behaving from $a^{-1}$ to $a^{-2}$ [17]. That is to say, less and less ballistic behaviour and at the end one has to obtain localization. Fig. 1b expresses this behaviour by showing that in very thin films smaller values of $p$ are needed to fit the data. These data are for

Au and they will be used when trying to understand quantized conductance experiments [18-24]. Notice the interesting point the 4K and the 300K data have the same resistivity for the very thin films, while there is a factor of four ratio for the thicker films of 77nm. This is a clear illustration of all the above discussion, the resistivity is determined by the thickness.

We proceed now by applying the above theory and experimental results to the conduction of electrons in narrow constrictions and nanocontacts [18-24]. Sharvin [25], following Knudsen, proposed that for the conduction problem corresponding to a diaphragm (see Fig.2a for an angle $\theta=\pi/2$ and $\delta\approx0$) when $a$ is much smaller than $l$, a ballistic resistance independent of $\rho$ and $l$ is obtained to be:

$$R_s = 4\rho l / 3A = (\frac{2e^2}{h} \frac{k_F^2 A}{4\pi})^{-1} \qquad (4)$$

where $A$ and $k_F$ are the area of the orifice and the Fermi wavevector of the electrons respectively, and $h$ and $e$ are the Planck constant and the electron charge. In the opposite limit, when $\kappa>>1$, one has the Maxwell result $\rho/a$. For the intermediate case, early work by Wexler [26] and recently Nikolic and Allen [27] showed that the resistance of the orifice for specular walls is:

$$R = R_s + R_{ohm} = R_s + \gamma(\kappa)\rho/a = R_s(1+\gamma(\kappa)a/l) \qquad (5)$$

where $\gamma(\kappa)$ is a factor of the order of unity, which is 1 and 0.67 when $\kappa$ tends to zero and infinity respectively. In addition, the papers of Ref. 26 and 27 prove that the two resistances, the ballistic and the ohmic one, are practically additives, an interesting result. It is clear then that when $a$ is much smaller than $l$ the Sharvin's resistance is dominant, i.e. the ballistic part dominates. *However, this is legitimate and valid in the case of a $\theta=\pi/2$ geometry for specular scattering in the walls, no roughness at the surface. Then one can see that $\rho$ behaves as $l^{-1}$.* But for smaller angles of $\theta$ this is not the case because the resistivities as well as the effective mean free path changes due to surface scattering with *the walls of the leads* which contact the orifice as has been demonstrated above by a vast set of theories and experiments in metallic films and wires.

Conductance quantization was studied using Landauer's formula [28-30] by using the geometry in Fig. 2 for $\theta=\pi/2$ and $\delta>\lambda$, where $\lambda$ is the Fermi wavelength of the electrons, and without roughness. It was calculated that the resistance is quantized in integral fractions units of $R_0=h/2e^2\approx 12900\Omega$. Experiments apparently showed the corresponding resistance plateaus [18-24] at the right "quantized" values in agreement with theory. However, one important point is that the data did not show the geometry of the metallic shape leading to "quantization". In fact only the experiments of Ohnishi et al [23] showed at the same time the values of resistance and the geometry of the constriction observed by TEM (Fig.2b) and the geometry was not $\theta\approx\pi/2$ but $\pi/4$. Later Torres et al [31] introduced the cone-like model of Fig.2 to discuss the problem and also find quantization if $\theta\leq\pi/4$. But it was studied without roughness (no paper in the field of conductance quantization is aware of the earlier work on increase of resistivity due to surfaced scattering); i.e. specular scattering in the cone walls. From our previous analysis, the metal surfaces have roughness and therefore the scattering is non-specular, with $p$ different of unity, in the model of Fuch-Sondheimer for Au, Na, Pb, Pt, and other metals. Therefore we would like to study the value of the $R_{Ohm}$ due to the rough walls of the cone leading to the contact constriction. We prove that this ohmic resistance for the experimentally observed values of $\theta\approx\pi/4$ [23] is of the same order of the Sharvin's resistance.

To perform this calculation we use the resistivity for a circular wire $\rho(r,p)$ as from the theory of Dingle (see ref. 5 AP formulae 28-31). Fig.3a shows the resistivities for different $p$ values as a function of the wire radius. Notice the growth of $\rho(r,p)$ behaving as $r^{-1}$ for small values of $r$. The resistance of the cone is obtained as a superposition of cylinders of radius $r$ and integrated from a given radius to $a/2$. Then we have:

$$R_{ohm} = \frac{\pi}{\alpha(\theta)} \int_{a/2}^{r_M} \frac{\rho(r)}{r^2} dr \qquad (6)$$

where $\alpha(\theta)$ is the cone solid angle.

This formula obtains the right result of the ohmic resistance for $\theta=\pi/2$ and it is very good up to $\pi/4$ which is the angle of the geometry in Fig.2b [23]. We have checked this formula against numerical simulations of the electric current density and field distributions using Maxwell equations and the software (FISINMAX). The comparison is presented in Fig.3b for different values of the parameters and there is agreement between formula 6 and the exact simulation.

To be more precise we have applied the theory to the experiments by Ohnishi et al [23], where the geometry is measured at the same time that the conductance is measured (Fig.2b). We have a value of the angle $\theta \approx \pi/4$. The calculated ohmic resistance is depicted in Fig.3b for $p=0.7$ to $0.3$ (these values are too large for very small values of $a$ and underestimate the ohmic resistance as observed from fitting data for films, see Fig.1). The $R_{Ohm}$ values obtained for one atom orifice $a \approx 0.3nm$ are, by substituting the values of $\rho_0$ as $2.2\ \mu\Omega.cm$ and $4.75 \mu\Omega.cm$ for Au and Na respectively, of the order of $10000\Omega$ and more, the claimed quantized resistance obtained in ref. 23 (Fig.2b) or in any other experiment on quantized conductance [18-24]. In addition, we would like to mention that the above result showed that the effective mean free path at the constriction is of the order of $a$, $R_{Ohm}$ behaves as $a^{-2}$. Therefore the Sharvin and ohmic resistances are of the same order and have an $a^{-2}$ behaviour (slope of the lines in Fig.3b); i.e. both resistances are indistinguishable. That is to say, the plateaus observed are a consequence of the addition of two resistances of the same order that are indistinguishable.

The results presented here have been ignored in all the previous work done in conductance quantization in metal nanocontacts and constriction. In fact ref. 24 is a review of the work and references as relevant for the problem as those quoted from 1-17 have not been mentioned in the review or in any other paper on conductance quantization. One of the authors in this paper (N.G.) is an active researcher in the field of conductance quantization and as everybody else was unaware of the work described here. We can forget this work, but this is to close the eyes to a fact, which has been well established and has relevance in the field of small or nanometer metallic structures, which is the work at hand. Therefore a reconsideration of the data on small nanocontacts seems necessary. On the other hand, we concluded that in order to have a negligible value of $R_{Ohm}$ one needs: *i)* that the surfaces of the walls of the leads are specular, which is complicated because not even the best grown films have this characteristic (real metal surfaces are rough), *ii)* that the two resistances of formula are not additives for non specular scattering but this sounds unphysical and is hard to prove. Notice that in the case we treat the resistance of the leads and the resistance of the contacts so that forcely they are additives, or *iii)* that the geometry forming the nanocontacts have $\theta \approx \pi/2$, which is also a peculiar results and in fact the only geometry we know in Fig.3b has $\theta \approx \pi/4$ with a large ohmic resistance. We have not considered roughness in the constriction of length δ, but it has been already discussed as another source of increase of the ohmic resistance [32], although it could be argued that for $δ ≈ λ$ this extra resistance could be overseen.

Finally, the effects described here although they may exist also in the 2DEG structures, the ohmic resistance may be much smaller because the structures are more perfect and have no roughness practically. But more because the geometry is the corresponding to a π/2 angle and the constrictions are not created by atoms but by an electrostatic repulsive field with negligible roughness, this makes a big difference. Although in any case deficiencies in conductance, even if small, should be observed.

**This work has been supported by the EU-FP6 BMR project.**

# Figure Captions

**Figure 1a** A summed up of measured thin film's resistivities vs thickness of several metals retrieved from corresponding references (-■- Pt [10], -●- Au (T≈0K) [11], -○- Au (T=250K) [11], -△- Bi [12], -☆-Cu [13], -□- Au [14], -◆- Ni-Fe Permalloy [15], -★- Al [16], -▲- CoSi$_2$ [17]). The resistivities are normalized by the metal bulk resistivity $\rho_0$ and the thickness is scaled by free mean path $l$ accordingly. The experimental data are fitted using the Fuchs-Sondheimer theory for the best values of $p$ as well as Mayadas and Shatzkes theory. The dashed line is for $\rho$ by M-S theory with $p=0.7$ and $R_{MS}=0.3$. The dotted line and solid line is for $\rho$ by F-S theory with $p=0.7$ and $0$ respectively.

**Figure 1b** The resistivities of Au film (thickness between 2 and 46 nm) at T≈0k (-●-) and T=250K (-■-) [11], and the fitting of them, using the same theory as in Fig.1a. The solid line is for $\rho$ by M-S theory with $p=0.1$ and $R_{MS}=0.15$, and the dashed line is for $\rho$ with $p=0.7$ and $R_{MS}=0.45$.

**Figure 2a** Profile of cone-likes nanocontact geometry, with *a and δ* is the narrowest width and the length of contact, *θ* is the half open angle and *r* is the radius of the cone.

**Figure 2b** TEM images of gold contacts, showing a half opening angle approximately π/4. Also the measured "quantized" resistance is presented for diffrents stages of the contact. These data are from Ref.23.

**Figure 3a** The resistivities for different *p* values as a function of the wire radius, based on the analysis from the theory of Dingle [7], see also Ref.5. The solid line, dashed line, the dotted line and the dash-dotted line represent for *p*=0, 0.3, 0.5 and 0.7 respectively.

**Figure 3b** The resistance of the cone integrated from a given diameter to *a* in Equ. (6), with θ=π/4 (dashed, dotted and solid lines of *p*=0.3, 0.5 and 0.7 respectively) and π/6 (dash-dotted line of *p*=0.7). Also the comparison with the accurate numerical simulations based on Maxwell equations for different values of *p* is presented. The lines of -○-, -□- and -△- are for θ=π/4, *p*=0.3, 0.5 and 0.7 respectively. The line with -■- is for θ=π/6 and *p*=0.7.

**FIGURES**

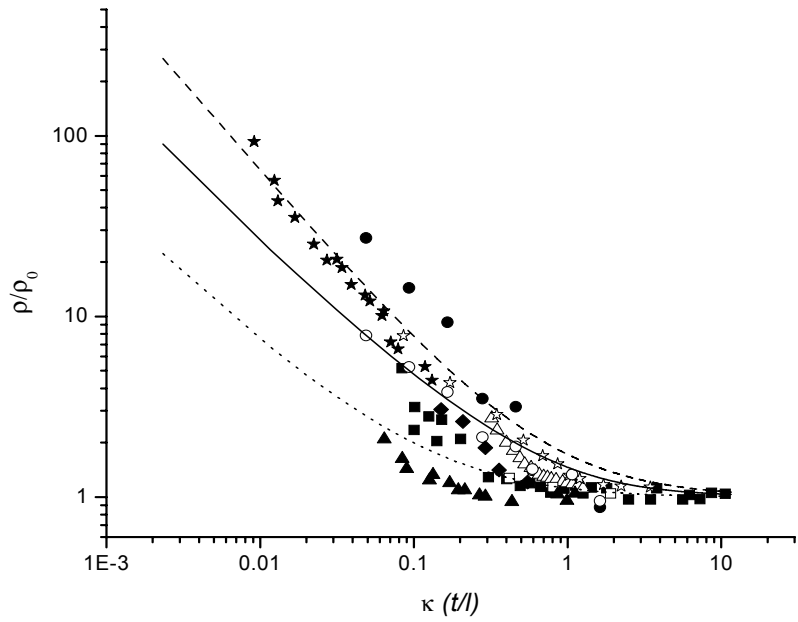

Figure 1a

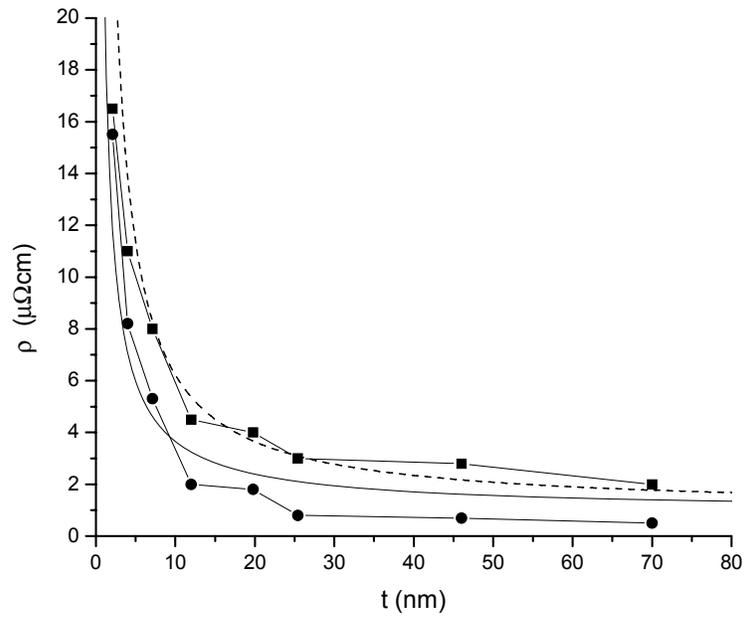

Figure 1b

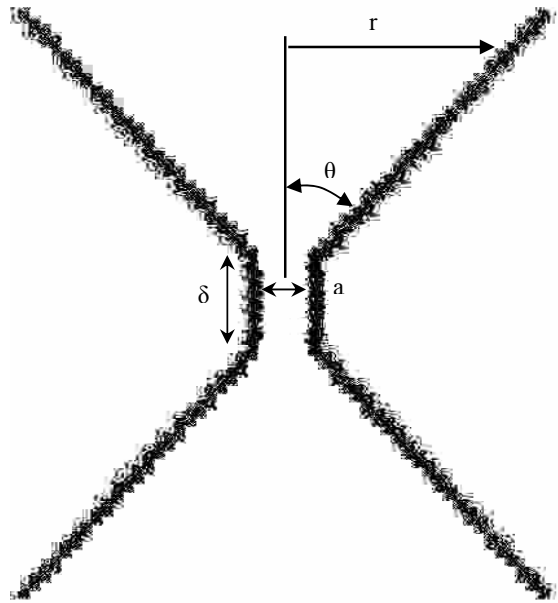

Figure 2a

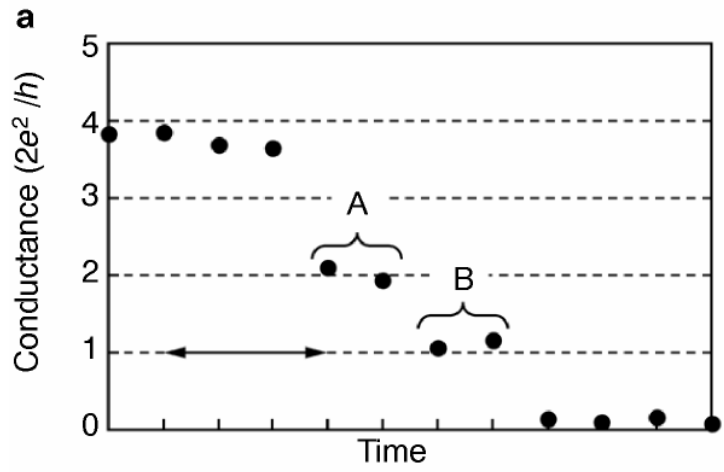

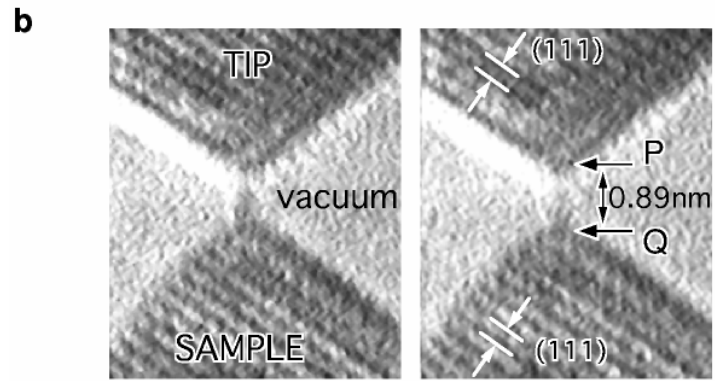

Figure 2b

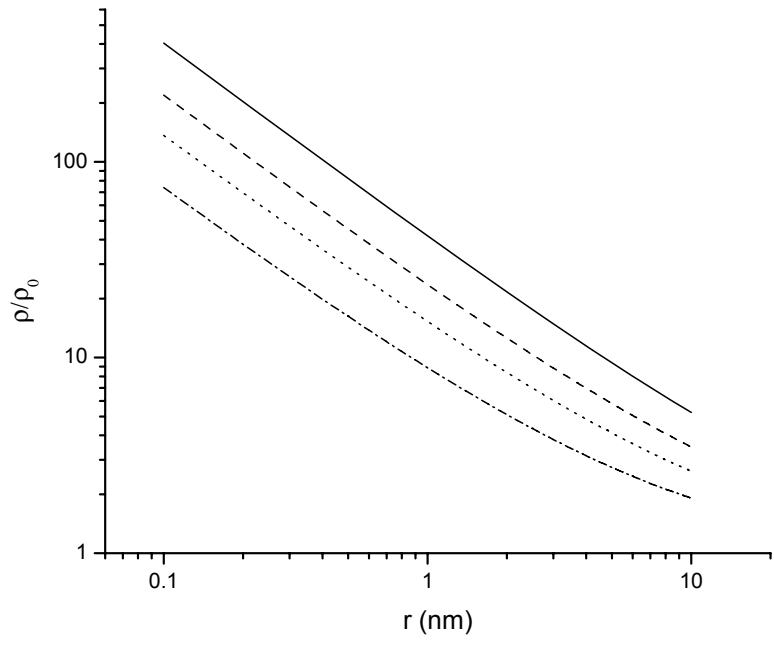

Figure 3a

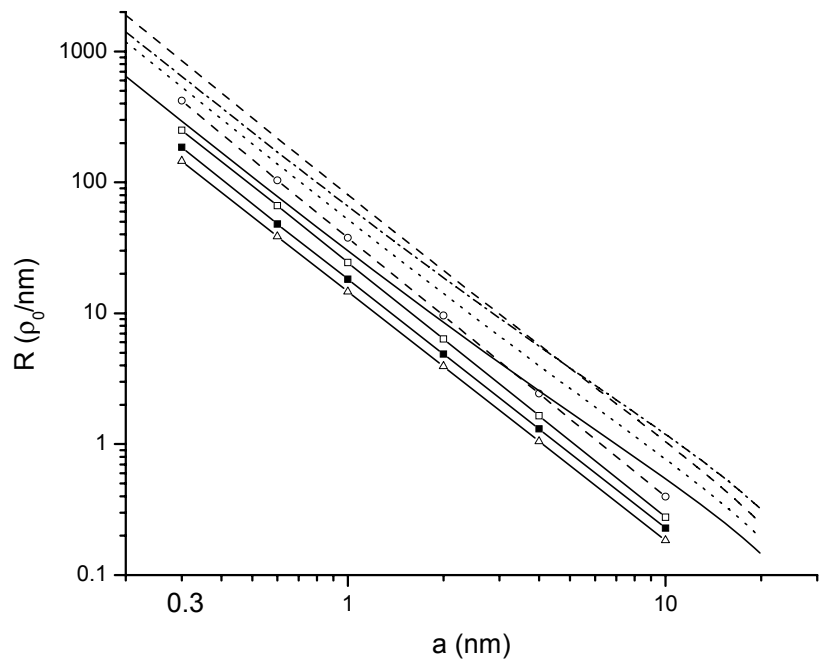

Figure 3b